\documentclass[12pt]{iopart}

\usepackage{epsfig}
\begin{document}

\title[Telescope of Geiger-M\"{u}ller counters]
{Educational studies of cosmic rays with telescope of Geiger-M\"{u}ller counters}

%\author{T.~Wibig$^1$, R.~Ko\l odziejczak$^{2,1}$, R.~Pierzy\'{n}ski$^{2,1}$,and %R.~Sobczak$^{2,1}$}
%\address{$^1$ A. So\l tan Institute for Nuclear Studies, Cosmic Ray Laboratory, POB-447
%90-950 \mbox{{\L }\'{o}d\'{z}-1,} Poland;}
%\address{$^2$ XII LO im. Stanis\l awa Wyspia\'{n}skiego, \L\'{o}d\'{z}, Poland;}
\author{T.~Wibig, K.~Ko\l odziejczak, R.~Pierzy\'{n}ski, R.~Sobczak}
\address{A. So\l tan Institute for Nuclear Studies, Cosmic Ray Laboratory, POB-447
90-950 \mbox{{\L }\'{o}d\'{z}-1,} Poland.}
%\address{$^2$ XII LO im. Stanis\l awa Wyspia\'{n}skiego, \L\'{o}d\'{z}, Poland;}
\ead{wibig@zpk.u.lodz.pl}
\begin{abstract}
A group of high school students (XII Liceum) in the framework of
the Roland Maze Project has built a compact telescope of three
Geiger-M\"{u}ller counters. The connection between the telescope
and PC computer was also created and programed by students
involved in the Project. This has allowed students to use their
equipment to perform serious scientific measurements concerning
the single cosmic ray muon flux at ground level and below. These
measurements were then analyzed with the programs based on the
'nowadays' knowledge on statistics. An overview of the apparatus,
methods and results were presented at several students conferences
and recently won the first prize in a national competition of high
school students scientific work. The telescope itself, in spite of
its 'scientific' purposes, is built in such a way that it is hung
on a wall in a school physics lab and counts muons continuously.
This can help to raise the interest for studying physics among
others. At present a few (3) groups of young participants of the
Roland Maze Project have already built their own telescopes for
their schools and some others are working on it. This work is a
perfect example of what can be done by  young people when
respective opportunities are created by more experienced
researchers and a little help and advice is given.
\end{abstract}

\maketitle

\section{Introduction}
\subsection{Cosmic rays}
Cosmic Rays (CR) were discovered in 1912 by Victor Hess (Nobel
Price in 1936). In his balloon flights he found an increase of the
discharge rate of an electroscope with the height of the balloon.
The nature of CR remained a mystery for quite a long time. At
present it is known that the radiation which we see deep in the
atmosphere is not the primary cosmic rays which enter from space.
The former consists mostly of stable nuclei (including single
protons), and also in small fraction electron and photons - the
only elementary particles which could survive the journey from CR
sources to the Earth (one can also mention neutrinos, but,
interacting only weakly, they can hardly be seen). Protons (and
neutrons trapped as nucleons in CR nuclei) interacting  kilometers
above ground with nuclei of the atmosphere create, if the energy
is high enough, short lived elementary particles - mostly pions.
These light mesons were discovered in the cosmic radiation by C.F.
Powell at Bristol and P.M.S. Blackett at Manchester in 1947.
Blackett was awarded the Nobel Physics Prize in 1948 and Powell in
1950.

Neutral pions decay almost immediately to high energy gamma quanta
(photons). High energy photons colliding with atomic nuclei can
create a pair of electrons and positrons. The positron is the
antiparticle of the electron and was predicted theoretically in
1928 by P.A.M.~Dirac (Nobel price in 1933). It was discovered in
cosmic rays by C.D.~Anderson in 1932 -- Nobel prize 1936 (shared
with V. Hess). Electrons and positrons interacting with charged
nuclei produced so-called brehmstrahlung photons, which in turn
can create  pairs etc. Sometimes, when the energy of the primary
CR particle is very high, electrons, positrons and photons in the
form of a cascade can reach ground level.

Charged pions can further interact strongly, creating several
generations of secondary particles, but sometimes, especially when
the density of nuclei to interact with is small (very high in the
atmosphere), they decay. One of the products of pion decay is the
muon. The muon is the lepton of the second generation in the
Standard Model, and as such it behaves exactly like an electron
The only difference is its mass. The muon is more massive than the
electron ($\sim$200 times). The muon was discovered in 1937 in
cosmic rays. A number of famous scientists had to work hard,
before eventually the true nature of the muon was established and
I.I.~Rabi could conclude the muon discovery with his famous "Who
ordered that?". C.D.~Anderson and S.H.~Neddermayer (1937),
J.C.~Street and E.C.~Stevenson (1937), F.~Rassetti, B.~Rossi and
N.~Nerson, Chaminade, A.~Freon and R.~Maze (1939), S.~Tomonaga and
G.~Araki (1940) all worked on the problem.

The ``Heavy electron'' -- the muon - interacts
electromagnetically, and because it is more massive, it is harder
to disturb its motion. When it is moving very fast, it is much
harder to force it to emit brehmstrahlung photons. A high energy
muon can easily travel not only the whole atmosphere, from the
point of origin to the Earth's surface, but also concrete walls
and floors. This property of the muon makes it the most abundant
particle in the cosmic ray flux at ground level. About 200 CR
particles cross each square meter area every second and about 3/4
of them are muons.

The electroscopes used in the pioneering years of cosmic ray
physics were soon replaced by more sophisticated equipment: cloud
chambers and Geiger-M\"{u}ller counters. The first allows one to
photograph tracks left by electrically charged particles in
overcooled clouds. The second gives an electric signal to any kind
of electronic register. The simplicity of the Geiger-M\"{u}ller
(GM) counter makes it very useful not only for scientific
purposes. The GM counters have been extensively used by astronauts
to explore other planets in early s.f. movies, and even by James
Bond in his first movie appearance in "Doctor No"\cite{bond}.

\begin{figure}
\centerline{\epsfig{file=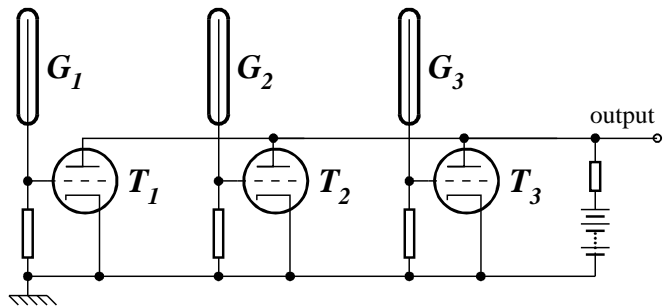,width=10cm}}
\caption{Three-fold coincidence circuit by B. Rossi.\label{fig1}}
\end{figure}

The GM counters used by B. Rossi around 1930 worked in a
coincidence mode made by three electronic valves\cite{rossi}. This
significant electronic achievement of its times is shown in
Fig.~\ref{fig1}.

\subsection{\bf The Roland Maze Project}

The mentioned cascade of elementary particles (photons, electrons,
positrons, muons, pions and other hadrons, e.g. kaons discovered
in 1947 by G. D. Rochester and C. C. Butler in cloud chamber
photographs of cosmic ray-induced events) reach the ground level
in the form of a thin (few meters) disc of a radius of hundreds of
meters. Such a phenomenon is called an Extensive Air Shower.
Extensive Air Showers (EAS) were seen for the first time by Roland
Maze with an apparatus built by him on the roof of the \`{E}cole
Normale Sup\'{e}rieure  in Paris. It consisted of three sets of GM
counters (each of area about 10$\times$10~cm$^2$) working in
different coincidence modes. In 1938, R.~Maze and P.~Auger
announced the discovery of huge cascades of charged particles
(EAS)\cite{eas}. The CR particles initiating such phenomena have
to have enormous energies, milions and even billions times greater
than `usual' resulting from radioactive decays and other known-in-
those days nuclear reactions.

Cosmic ray studies have been recently intensively developed,
especially in the region of the upper limit of the energy
spectrum, where several events involving cosmic ray particles with
energy exceeding 10$^{20}$~eV ($\approx$~50~J) have been observed.
Their existence is a very serious physical problem. It was number
4 on the additional problem list published in 2005 in the Science
magazine \cite{science}: {\it Where do ultrahigh-energy cosmic
rays come from? Above a certain energy, cosmic rays don't travel
very far before being destroyed. So why are cosmic-ray hunters
spotting such rays with no obvious source within our galaxy?}

The experimental setup for recording such events usually consists
of a number of relatively simple particle detectors spread over a
large area. Nowadays, the single detector is a scintillation (or
Cerenkov) counter connected to the electronic system with a
coincidence trigger and converters of time and amplitude of a
signals to digital codes. The essence of large area experiments is
a method of synchronization and communication between detectors
and a system of collecting and storing the data.

This is the point where high school education can meet high
science. It is hard to imagine another subject of such great
importance which can be studied jointly by scientists and
students. It is not surprising that at present there are related
projects under constructions in the USA \cite{inneUSA} and in
Europe\cite{inneEU}.

One of them is the Roland Maze Project\cite{maze}. The proposed
EAS detection stations would be placed in the buildings of high
schools.

The detection system of one station (school) allows one to conduct
(in parallel with the main scientific object of the project:
studies of extremely high energy cosmic rays) independent
observations and studies for each group participating in the
project.  It covers the whole, wide region of cosmic ray particle
energies, giving the ability to study geophysics and atmospheric
phenomena as well as monitoring the Sun's activity and space
weather from one side, up to the properties of typical EAS on the
other.

We have gathered many students interested in making "big physics".
The constructions of the project detectors and all the systems is
going on, but in spite of that in order not to lose the initial
impact we have proposed to the students many other activities.

\section{Telescope of Geiger-M\"{u}ller counters}

One of the many byproducts of the Roland Maze Project is the idea
of making small cosmic ray detectors/counters which can be hung on
the school wall showing to everybody that cosmic rays are
everywhere and raising the interest in science in general.

The counters are arranged in telescopes of three Geiger-M\"{u}ller
counters working in coincidence. Such a setup reminds us of the
very first array of Maze.

\begin{figure}
\centerline{\epsfig{file=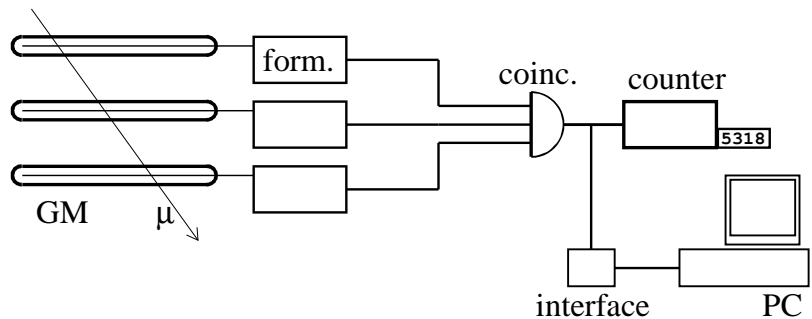,width=12cm}}
\caption{The schema of the telescope of GM counters.\label{fig2}}
\end{figure}

The telescopes are to be built entirely by groups of students from
each school. The only parts which can't be made by them are GM
counters, which are made in the A. So\l tan Institute for Nuclear
Studies in {\L }\'{o}d\'{z}. The counters are of glass with
external cathodes. This kind of GM counters is called the Maze
type\cite{mtype}. The electronic schemes of the particular
circuits are also to some extent created by the students,
especially if there is one in the group with some electronic
experience. If there is none, a general scheme is given and the
particular solutions are then found empirically. This leads to
breaking some electronic components, but eventually it leads to
great satisfaction which is one of the more important factors when
doing science and what is hard to explain to the students in any
other way.

\subsection{A Particular solution}

The telescope from which results we want to present in this paper
was made by students of the XII Liceum in {\L }\'{o}d\'{z}.

%\begin{figure}
%\centerline{\epsfig{file=schemgm.eps,width=12cm}}
%\caption{The schema of the telescope of GM counters.\label{fig2}}
%\end{figure}

A schematic view of the telescope is shown in Fig.~\ref{fig2}.

The high voltage of about 1500~V needed to supply the GM tubes is
created by a modified TV HV transformer with a primary winding
connected to a simple pulse generator of 5-12~V and secondary to
the Villard cascade. The coincidence was realized with standard
TTL monostables with duration time of about 2~$\mu$s. The 4-digits
7 segment display of about 1~inch height was used to show the
number of counts.

The telescope was equipped with a simple interface built on a base
of the 555 circuit used to connect it to the PC class computer.
The interface was originally programed under DOS in BASIC. The
computer is able to continuously register telescope counts and to
write them successively on the computer disk.

\section{Results}
\subsection{Stability}

\begin{figure}
\centerline{\epsfig{file=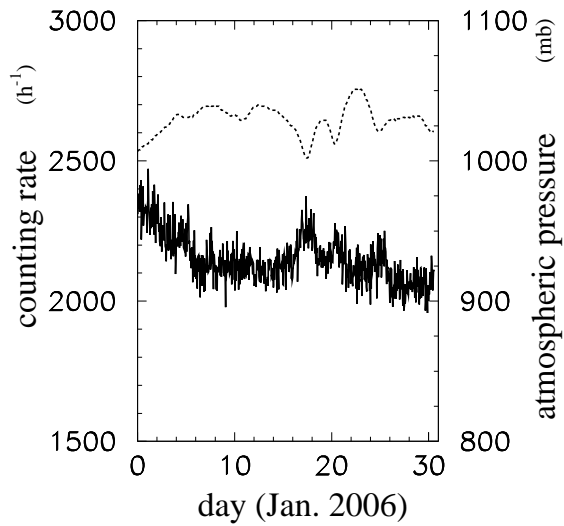,width=8cm}}
\caption{The telescope counting ratio (solid line and left scale) and atmospheric
pressure (dashed line and right scale) registered in January 2006.\label{fig3}}
\end{figure}

In  Fig.~\ref{fig3} we show the number of counts registered every
hour since the beginning of  2006  to the end of January. As  is
seen there are no abrupt (unexpected, as will be discussed later)
changes, and this causes us to state that the telescope is working
properly and rather stable.

\subsection{Barometric coefficient}
The single muon flux changes with time, as has been known for a
long time. The most pronounced modulation relates to the
atmospheric pressure. Low energy ($\sim$GeV) muons originate in
the upper levels of the atmosphere and have to traverse almost the
whole atmosphere (hundreds grams per cm$^2$ of air). The amount of
the air above us changes continuously. These changes are the
subject of interest to billions of people around the world. They
are presented several times every day in TV weather reports on
most of the TV channels as values of the atmospheric pressure. The
pressure of 1000 mb informs us that above every square centimeter
there is about 1 kg of air.

If the atmospheric layer to traverse is thicker then the flux of
muons is diminished, due mainly to energy losses. Thus an
anticorrelation of telescope counting rate and atmospheric
pressure is expected. The values of the pressure from the local
meteorological station Lublinek were taken from the respective web
page and they are plotted in Fig.~\ref{fig3}. Fortunately,
according the very rapid and substantial changes of the pressure
in January, the anticorrelation is clearly seen by the naked eye.

\begin{figure}
\centerline{\epsfig{file=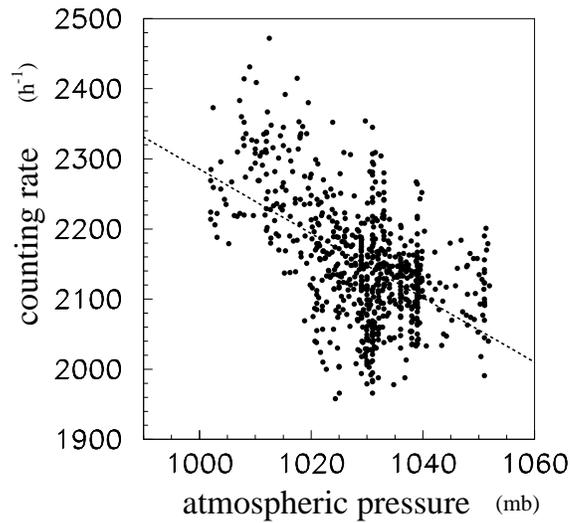,width=8cm}}
\caption{Scattered plot telescope counting rate vs. atmospheric pressure registered in January 2005. Each point represents one hour counting results. \label{fig4}}
\end{figure}

To study it in detail in Fig.~\ref{fig4} we show the scatter plot
of telescope counting rate vs. atmospheric pressure. The dashed
line plotted there is the best linear fit. To relate our result
with the one known from literature we can expressed it as a
relative change of the counting rate with respect to the increase
of the pressure of 1~mb. Such a value is called {\em barometric
coefficient}. Our result is -0.21$\pm$0.04\%/mb.

\subsection{Muon absorption in the ground}

\begin{figure}
\centerline{\epsfig{file=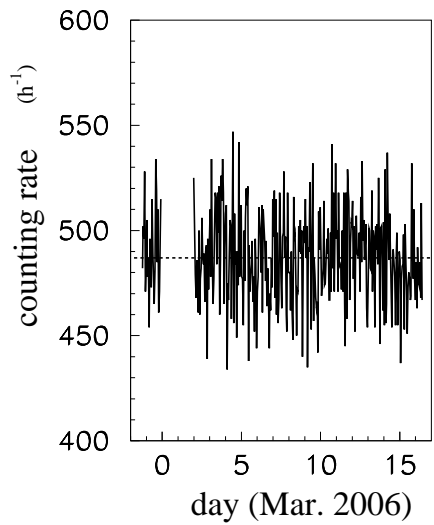,width=8cm}}
\caption{Telescope counting rate underground. \label{fig5}}
\end{figure}

The \L \'{o}d\'{z} EAS array of the A. So\l tan Institute for
Nuclear Studies has its detection point in the underground
laboratory placed 15~m below  ground level.  At present it is used
as the site of the prototype of the \L \'{o}d\'{z} ''space weather
station''. The students' GM telescope was placed there and it
measured the single muon flux for a few days. The results are
shown in Fig.~\ref{fig5}

Calculations show that muons which are able to reach the
underground lab have to have energy greater than about 5~GeV. The
measured decrease of the muon flux agrees with the value expected
from the known integral energy spectrum.

\section{Discussion}
\subsection{The Main results}
The telescope made by the high school students, operating for one
month, produced data which allowed students to determine precisely
the value of the barometric coefficient as -0.21$\pm$0.04\%/mb.
This value is just what can be expected from the literature (-0.18
to -0.20\%/mb.) confirming the principles of the methods and
solutions used.

It is interesting to mention the work of an Italian group
Ref.\cite{wlosi}. The general idea was similar - to study the
barometric effect with the help of simple computerized apparatus
based on one small GM counter. Comparison shows that the results
obtained by them -0.051$\pm$0.015 (or even -0.023$\pm$0.009, based
on 3.5 months of running the experiment) is significantly
different from ours. The simple explanation of this difference is
that the coincidence of three GM counters used by us warrant that
the counts we registered are definitively particles crossing areas
of our GMs, and taking into account the thickness of the GM
counter walls they have to be mostly cosmic ray generated muons.
Without such a coincidence the single GM counter is measuring any
kind of radiation which can penetrate the active GM counter
volume. Only a fraction of it depends on atmospheric pressure.

The known flux of vertical muons given by PDG \cite{PDG} gives the
expected counting rate, and after simple geometrical integration
we checked that it is consistent with our measured value (shown in
Fig.~\ref{fig2}).

\subsection{Significant achievements}
The value of the barometric coefficient itself, in spite of its
correctness, is not the most important achievement of the present
work. Several points have to be emphasized:
\begin{itemize}
\item[-] students (K.K., R.P., and R.S.) designed the scientific
instrument; \item[-] students built it, checking, correcting and
testing every part of it; \item[-] students assembled all the
parts together making the fully operational apparatus; \item[-]
students performed the measurements; \item[-] students wrote the
analysis programs and analyzed the data; \item[-] students
presented their achievements on several occasions (starting from a
presentation given by them in a forum of their own class,
presentation in Lodz high school competition on physics,
presentation on "3$^{rm rd}$ Roland Maze Scientific Session", and
recently in the competition of national scientific works for high
school students organized by the Polish Academy of Science and
INS, where they won The First Prize).
\end{itemize}
The role of more experienced scientists (e.g. T.W.) participating
in the telescope work should also be expressed precisely to avoid
any suspicions (as we noticed on a few occasions). First and
obvious is to formulate the subject of investigation. Then the
subject has to be introduced to the students in the way attractive
enough to mobilize young people to get the job. Some help in
electronics was of course needed. Specific questions concerning
usage of elements to build up properly working circuits can be
mostly answered with the help of the internet. Some personal help
and assistance was necessary, of course, when testing built parts
and assembling the telescope.

The construction process takes time, of course. For this
particular telescope this time was approximately one year.
Students worked on the telescope in their free time, mainly on
Saturdays, when they  spent their time without any collision with
school duties. During this year students gathered not only
abilities of making electronic devices, but in the meantime they
improved their knowledge also of physics and of cosmic rays in
particular. To achieve this the assistance of a more experience
scientist was important, of course.

When the telescope was built, a series of measurements was
performed. The idea of what to measure comes naturally. The test
of stability is obvious. The experimental setup produced a number
of files with huge (on the respective scale) amounts of data. The
interface and registration program was created in such a way that
each single registration was written on to the disk. The help of
the scientist was then necessary to draw the attention of students
to the smooth changes of the counting rate seen in
Fig.~\ref{fig2}. This was also a good occasion on which to study
the foundations of statistics, e.g., the variation of the Poisson
distributed random variable. The explanation of these changes as
due to the barometric effect also had to be introduced to the
students by T.W. Then, after a short lesson on statistics
(straight line fitting, $\chi^2$ methods, e.t.c.), the students
wrote their first programs in C and ran them to get the result
presented above.

\subsection{Further planned measurements}
The telescope can be used also to make other measurements among
which the most obvious is to determine the zenith angle dependence
of single cosmic ray muons. Studies of the temperature effect as
well as searching for periodic muon flux variations, 27 days, one
day, semidiurnal to name only the shortest) are possible subjects
of further interesting studies. The attractive subject is to
looking for a correlations of CR flux with other phenomena, to
mention only the famous examples (?) in Janathan Swift's {\it
Gulliver's Travels} (almost 300 years ago), and William Herschel's
studies from very beginning of the XIX century ( about 200 years
ago), and more recently (2003) studies on the influence of the
cosmic ray intensity on the wheat price in medieval England
\cite{england}. The question if CR muons are correlated with
average marks in school tests, can attract large numbers of
students.

\section{Conclusions}
We have shown that cosmic rays are  one of the subjects of
contemporary physics which are very useful to raise the interest
of science amongst students of high schools. The present work
proves that they are able to construct the apparatus which may be
used to give quite accurate data on cosmic rays at ground level.
The value of the barometric coefficient can be obtained and other
interesting studies can be performed. The analysis of the data
gives a perfect possibility for students to be introduced to
statistical methods at a level not available in standard courses.

The Geiger-M\"{u}ller telescope hanging on a classroom wall and
showing continuously the number of mouns crossing it works well at
increasing the horizons of mind not only of young people but of
all who see it.


\section*{References}
\begin{thebibliography}{10}
\bibitem{bond}
\verb+http:\\www.jamesbondmm.co.uk/q-branch/drno-gadgets.php+.
\bibitem{rossi}  B. Rossi, {\it Cosmic Rays}, McGraw-Hill, New York, (1964).
\bibitem{eas} P. Auger, R. Maze, T. Grivet-Meyer, {\em Comptes rendus}, {\bf 206},
1721 (1938).
\bibitem{science}
D. Kennedy and C. Norman, Science {\bf 309}, 78 (2005);
 \verb+http:\\www.sciencemag.org/sciext/125th/#online+.
\bibitem{inneUSA}
\verb+http:\\csr.phys.ualberta.ca/~alta+, \\
\verb+http:\\www.phys.washington.edu/~walta+,\\
\verb+http:\\www.chicos.caltech.edu+, \\
\verb+http:\\physics.unl.edu+, \\
\verb+http:\\faculty.washington.edu/~wilkes/salta+, \\
 \verb+http:\\www.phyast.pitt.edu/~jth/CosRayHS.html+,\\
\verb+http:\\www.hep.physics.neu.edu/scrod+, \\
\verb+http:\\www.phy.bnl.gov/~takai/MariachiWeb+.
\bibitem{inneEU}
\verb+http:\\www.hisparc.nl+, \\
\verb+http:\\www.hisparc.nl/eurocosmics.php+, \\
\verb+http:\\hep.ph.liv.ac.uk/~green/cosmic/home.html+, \\
\verb+http:\\www.particle.kth.se/SEASA+.
\bibitem{maze}
J.~Gawin {\it et al.}, %The Roland Maze Project ,
Acta Physica Polonica {\bf B33}, 349 (2002);
\verb+http:\\maze.u.lodz.pl+.

\bibitem{mtype}
S.~Michalak, B.~Mowczan and A.~Zawadzki,
%?L.Allongement du paliers des compteurs GM a cathode en %verre?,
Acta Phys. Pol. {\bf 13}, 145 (1954).
\bibitem{wlosi}
B. Famoso, P. La Rocca and F. Riggi ,%? An educational study of the barometric effect of cosmic %rays with a Geiger counter?,
Phys. Educ. {\bf 40}, 461 (2005).
\bibitem{PDG}
S. Eidelman {\it et al.},
Phys. Lett. B 592, 1 (2004); (URL: \verb+%http:\\pdg.lbl.gov/+).
% and 2005 partial update for the 2006 edition available on the PDG WWW pages
\bibitem{england}L.A. Pustilnik, L.I.~Dorman, and G.~Yom~Din
Proc. 28$^{\rm th}$ Intl.  Cosmic Ray Conf., Tsukuba, 4131 (2003).
\end{thebibliography}
\end{document}